\documentclass[%
 aip,
 jap,%
 amsmath,amssymb,
 reprint,%
]{revtex4-1}

\usepackage{graphicx}
\usepackage{epstopdf}
\usepackage{dcolumn}
\usepackage{bm}

\begin{document}

\preprint{AIP/123-QED}

\title[Faraday Rotation and One-Way Waves]{Faraday Rotation, Band Splitting, and One-Way Propagation of Plasmon Waves on a  Nanoparticle Chain}

\author{N. A. Pike}
 \affiliation{D\'{e}partment de Physique, Universit\`{e} de Li\'{e}ge, 4000 Sart Tilman, Belgium}
\affiliation{ Department of Physics, The Ohio State University, Columbus, Ohio 43210, USA }%
\author{D. Stroud}%
 \email{Stroud@physics.osu.edu}
\affiliation{ Department of Physics, The Ohio State University, Columbus, Ohio 43210, USA }%

\date{\today}

\begin{abstract}
We calculate the dispersion relations of plasmonic waves propagating along a chain of semiconducting or metallic nanoparticles in the presence of both a static magnetic field ${\bf  B}$ and a liquid crystalline host. The dispersion relations are obtained using the quasistatic approximation and a dipole-dipole  approximation to treat the interaction between surface plasmons on different nanoparticles.  For a plasmons propagating along a particle chain in a nematic liquid crystalline host with both ${\bf B}$ and the director parallel to the chain, we find a small, but finite, Faraday rotation angle. For ${\bf B}$ perpendicular to the chain, but director still parallel to the chain, the field couples the longitudinal and one of the two transverse plasmonic branches.  This coupling is shown to split the two branches at the zero field crossing by an amount proportional to $|{\bf B}|$.  In a cholesteric liquid crystal host and an applied magnetic field parallel to the chain, the dispersion relations for left- and right-moving waves are found to be different.  For some frequencies, the plasmonic wave propagates only in one of the two directions.
\end{abstract}


\maketitle
Ordered arrays of metal particles in dielectric hosts have many remarkable properties~\cite{meltzer,maier2,tang,park,pike2013,pike2013a}. For example, they support propagating modes which are linear superpositions of so-called "surface" or "particle" plasmons.   In dilute suspensions of such nanoparticles, these surface plasmons give rise to characteristic absorption peaks, in the near infrared or visible, which play an important role in their optical response, and which have recently been observed in semiconductor nanoparticles as well as metallic ones~\cite{Faucheaux,Hsu}.  For ordered chains, if both the particle dimensions and the interparticle separation are much smaller than the wavelength of light, one can readily calculate the dispersion relations for both transverse ($T$) and longitudinal ($L$) waves propagating along the chain, using the quasistatic approximation, in which the curl of the electric field is neglected.

In a previous paper, we calculated these dispersion relations for metallic chains immersed in an anisotropic host, such as a nematic or cholesteric liquid crystal (NLC or CLC)~\cite{pike2013}.   Here, we consider the additional effects of a static magnetic field applied either parallel and perpendicular to a chain of nanoparticles.  In order to obtain a larger effect from the magnetic field, we will also consider doped semiconducting nanoparticles.  Such nanoparticles have a much lower electron density than typical metallic nanoparticles.  For example, the electron density in semiconductor nanoparticles, such as the Cu$_{2-x}S$ nanoparticles whose optical properties have recently been studied~\cite{Faucheaux}, can be adjusted over a broad range from  $10^{17} - 10^{22}$ cm$^{-3}$ or even lower.   The largest effects are obtained with electron densities towards the lower end of this range.   We find, for a parallel magnetic field orientation, that a linearly polarized $T$ wave undergoes a Faraday rotation as it propagates along the chain.  For a field of $2$ Tesla and a suitably low electron density, this Faraday rotation  can be at least 1 degree per ten interparticle spacings.  In this case, for the  parallel field orientation, the NLC quantitatively modifies the amount of Faraday rotation, but there would still be rotation without the NLC host.

We also consider the  propagation of plasmonic waves along a nanoparticle chain but with a cholesteric liquid crystal (CLC) host . In this case, if the magnetic field is parallel to the chain and the director rotates about the chain axis with a finite pitch angle, we show that the frequencies of left- and right-propagating waves are, in general, not equal. Because of this difference, it is possible, in principle, that for certain frequencies, a linearly polarized wave can propagate along the chain only in one of the two possible directions.  Indeed, for sufficiently low electron concentration, we do find one-way propagation in certain frequency ranges.   This realization of one-way propagation is quite different from other proposals for one-wave waveguiding~\cite{yu,mazor,hadad,wang,dixit}.

The remainder of this article is organized as follows:  First, we use the formalism of Ref.~\citenum{pike2013} to determine the dispersion relations for the $L$ and $T$ waves in the presence of an anisotropic host and a static  magnetic field.   Next, we give simple numerical examples and finally we provide a brief concluding discussion. 

\section{Formalism}
We consider a chain of identical metallic or semiconducting nanoparticles, each a sphere of radius $a$, arranged in a one-dimensional periodic lattice along the $z$ axis, with lattice spacing $d$, so that the $n^{th}$ particle is assumed centered at $(0, 0, nd)$ ($-\infty < n < + \infty$).   The propagation of plasmonic waves along such a chain of nanoparticles has already been considered extensively for the case of isotropic metal particles embedded in a homogeneous, isotropic medium~\cite{brong}.  In the present work, we calculate, within the quasistatic approximation,  how the plasmon dispersion relations are modified when the particle chain is immersed in both an anisotropic dielectric, such as an NLC or CLC, and a static magnetic field. We thus generalize earlier work in which an anisotropic host is considered without the magnetic field~\cite{pike2013a,pike2013}.

In the absence of a magnetic field, the medium inside the particles is assumed to have a scalar dielectric function.  If there is a magnetic field ${\bf B}$ parallel to the chain (which we take to lie  along the $z$ axis), the dielectric function of the particles becomes a tensor, $\hat{\boldsymbol \epsilon}$.   In the Drude approximation, the diagonal components are $\epsilon(\omega)$,  while $\epsilon_{xy} = -\epsilon_{yx}= iA(\omega)$ and all other components vanish.  In this case, the components of the dielectric tensor take the form
\begin{equation}
\epsilon(\omega) = 1 - \frac{\omega_p^2}{\omega(\omega+ i/\tau )} \rightarrow  1 - \frac{\omega_p^2}{\omega^2},
\label{eq:epsw}
\end{equation}
and
\begin{equation}
A(\omega) = -\frac{\omega_p^2\tau}{\omega}\frac{\omega_c\tau}{(1-i\omega\tau)^2} \rightarrow \frac{\omega_p^2\omega_c}{\omega^3},
\label{eq:aw}
\end{equation}
where $\omega_p$ is the plasma frequency, $\tau$ is a relaxation time, and $\omega_c$ is the cyclotron frequency, and the second limit  is applicable when $\omega\tau \rightarrow \infty$.  We will use Gaussian units throughout.   While this approximation may be somewhat crude, especially for semiconducting nanoparticles, it should be a reasonable first approximation.
\begin{figure}[t]
\includegraphics[width=0.45\textwidth]{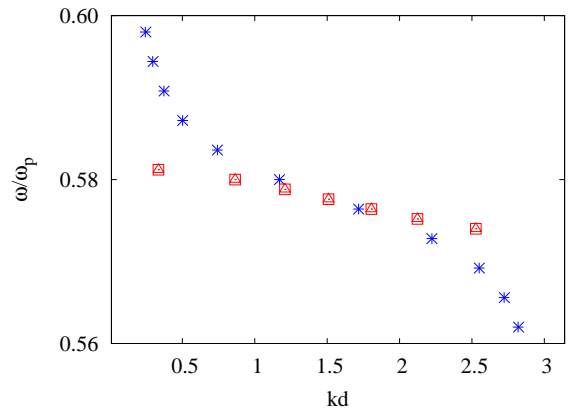} 
\caption{Blue symbols (x's and +'s): Dispersion relations for left (x) and right (+) circularly polarized $T$  plasmon waves propagating along a chain of  nanoparticles immersed in a NLC with both the director and a magnetic field parallel to a chain. The particles are described by a Drude dielectric function with $\omega_p \tau = 100$ and $\omega_c/\omega_p = 0.07$. Red symbols (open squares and triangles):  Same as the blue symbols, but assuming no single-particle damping ($\omega_p\tau = \infty$).   In both cases, the splitting between left and right circularly polarized waves is not visible on the scale of the figure (but the rotation is visible in Fig.\ 2). For $\omega_p = 5.0\times 10^{12}$ sec$^{-1}$, the chosen $\omega_c/\omega_p$ corresponds to $B \sim 2$ Tesla.}
\label{figure1}
\end{figure}

The dielectric function of the liquid crystal host, for either the NLC or CLC case, is taken to be that described in Ref.~\citenum{pike2013}.  The dispersion relations for the surface plasmon waves are determined within the formalism of Ref.~\citenum{pike2013}.   Specifically, we write down a set of self-consistent equations for the coupled dipole moments; these are given in Ref.~\citenum{pike2013} as Eq. (9), and repeated here for reference:
\begin{equation}
{\bf p}_n = -\frac{4\pi a^3}{3}\hat{\bf t}\sum_{n^\prime \neq n}\hat{\cal G}({\bf x}_n - {\bf x}_{n^\prime}){\bf p}_{n^\prime}.
\label{eq:selfconsist}
\end{equation}
Here 
\begin{equation}
\hat{\bf t} = \delta\hat{\boldsymbol{\epsilon}}\left(\hat{\bf 1}-\hat{\bf \Gamma}\delta\hat{\boldsymbol{\epsilon}}\right)^{-1}
\label{eq:tmatrix}
\end{equation}
is a ``t-matrix'' describing the scattering properties of the nanoparticle spheres in the surrounding material, $\hat{\cal G}$ and $\hat{\bf \Gamma}$ are a $3\times 3$ Green's function and depolarization matrix given in Ref.~\citenum{pike2013}, $\hat{\bf 1}$ is the $3\times3$ identity matrix, and $\delta \hat{\boldsymbol{\epsilon}} = \hat{\boldsymbol{\epsilon}}- \hat{\boldsymbol{\epsilon}}_h$,  where $\hat{\boldsymbol \epsilon}_h$ is the dielectric tensor of the liquid crystal host.   

\begin{figure}[t]
 \centering
 \includegraphics[width=0.45\textwidth]{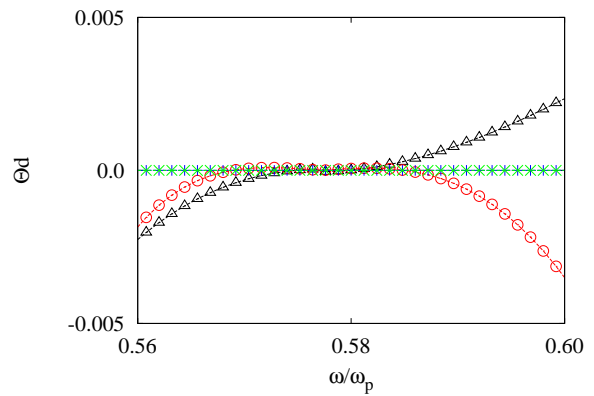} 
 \caption{Real and imaginary parts of $\theta d$, the rotation angle per interparticle spacing (in radians), as a function of frequency, assuming $\omega_c/\omega_p = 0.07$.  Blue $+$'s (real part) and green $x$'s (imaginary part of $\theta d$):  Drude model with no damping ($\omega_p\tau \rightarrow \infty$).   Black triangles (real part) and red  circles (imaginary part of $\theta d$):  Drude model with finite damping ($\omega_p\tau = 100$).    In both cases, the magnetic field and the director of the NLC are assumed parallel to the chain axis, as in Fig.~\ref{figure1}. The dotted lines merely connect the points.}
 \label{figure2}
\end{figure}

\subsection{Nematic Liquid Crystal}

We first consider a chain of such particles placed in an NLC host with ${\bf B}\| \hat{z}$ and parallel to the liquid crystal director $\hat{n}$.   Using the formalism of Ref.~\citenum{pike2013}, combined with Eq.~\eqref{eq:selfconsist}, we obtain two coupled sets of linear equations for the transverse ($T$) components of the polarization, $p_{nx}$ and $p_{ny}$.   The solutions are found to be left- and right-circularly polarized transverse waves with frequency $\omega$ and wave number $k_{\pm}$, where the frequencies and wave numbers are connected by the dispersion relations in the nearest-neighbor approximation 
\begin{equation}\label{eq:nlc_parallel_disp}
1= -\frac{2}{3} \frac{a^3}{d^3} \frac{\epsilon_\|}{\epsilon_\perp^2}\left(\frac{\epsilon(\omega)-1}{\epsilon(\omega)+2} \mp \frac{3 A(\omega)}{(\epsilon(\omega)+2)^2}\right)\cos(k_\pm d),
\end{equation}
where we use the notation of Ref.~\citenum{pike2013}.  These equations are accurate to first order in $A(\omega)$.  The longitudinal ($L$ or  $z$) mode is unaffected by the magnetic field.  Since the frequency-dependences of both $\epsilon(\omega)$ and $A(\omega)$ are assumed known, these equations represent implicit relations between $\omega$ and $k_\pm$ for these $T$  waves.

If ${\bf B} \|  \hat{x}$ while both $\hat{n}$ and the chain of particles are parallel to $\hat{z}$, then the $y$ and $z$ polarized waves are coupled   The dispersion relations are obtained  as solutions to the coupled equations 
\begin{eqnarray}
p_{0y} & = &  \frac{-2a^3}{3d^3}\left[\frac{\epsilon_\|}{\epsilon_\perp^2}t_{yy}p_{0y} - \frac{2}{\epsilon_\perp}t_{yz}p_{0z}\right]\cos (kd), \nonumber \\
p_{0z} & = & \frac{-2a^3}{3d^3}\left[-\frac{\epsilon_\|}{\epsilon_\perp^2}t_{yz}p_{0y} -\frac{2}{\epsilon_\perp}t_{zz}p_{0z}\right]\cos (kd).
\label{eq:rotyz2}
\end{eqnarray}
These $y$ and $z$ modes are uncoupled from the $x$ modes. 

If we solve this pair of equations for $p_{0y}$ and $p_{0z}$ for a given $k$, we obtain a nonzero solution only if the determinant of the matrix of coefficients vanishes. For a given real frequency $\omega$, there will, in general, be two solutions for $k(\omega)$ which decay in the $+z$ direction.   These correspond to two branches of propagating plasmon (or plasmon polariton) waves, with dispersion relations which we may write as $k_\pm(\omega)$.   The frequency dependence appears in $t_{yz}$, $t_{yy}$, and $t_{zz}$, which depend on $\omega$ [through $\epsilon(\omega)$ and $A(\omega)$].  However, unlike the case where the magnetic field is parallel to the  $z$ axis, the waves are elliptically rather than circularly polarized.

\subsection{Cholesteric Liquid Crystal}

We now consider immersing the chain of semiconducting nanoparticles in a CLC in the presence of a static magnetic field with ${\bf B}\| \hat{z}$ and the chain.  A CLC can be thought of as an NLC whose director is perpendicular to a rotation axis (which we take to be $\hat{z}$), and which spirals about that axis with a pitch angle $\alpha$ per interparticle spacing.  For a CLC, if we include only interactions between nearest-neighbor dipoles, the coupled dipole equation [Eq.~\eqref{eq:selfconsist}] takes the form
\begin{equation}
{\tilde p}_n = -\frac{4\pi a^3}{3}[\hat{\bf R}^{-1}(z_1)\hat{\bf t}\hat{\cal G}\cdot{\tilde p}_{n+1} +\hat{\bf R}(z_1)\hat{\bf t}\hat{\cal G}\cdot{\tilde p}_{n-1}], 
\label{eq:selfcon_two}
\end{equation}
as is shown in Refs.~\citenum{pike2013} and~\citenum{pike2013a}.  Here $\tilde{p}_n = {\bf R}_n(z)p_n$ and ${\bf R}_n(z)$ is a $ 2\times2 $ rotation matrix for the director $\hat{n}(z)$. If ${\bf B} \| \hat{z}$, the two $T$ branches are coupled.   One can write a  $2\times 2$ matrix equation for the coupled dipole equations in the rotated  ${x}$ and ${y}$ directions.  This equation is found to be
\begin{equation}
\tilde{\bf  p}_0 = -\frac{2 a^3}{3 d^3}\hat{\bf M}(k, \omega)\cdot \tilde{\bf p}_0,
\label{eq:disperchiral}
\end{equation}
where $\tilde{\bf p}_0$ is the rotated two-component column vector whose components are $\tilde{\bf p}_{x0}$ and $\tilde{\bf p}_{y0}$.   The components of the matrix $\hat{\bf M}(k,\omega)$ are found to be
\begin{eqnarray}
M_{xx} &=& \epsilon_1\lbrack t_{xx}\cos(kd)\cos(\alpha d) + it_{xy}\sin(kd)\sin(\alpha d)\rbrack \nonumber \\
M_{yy} &=& \epsilon_2\lbrack t_{yy}\cos(kd)\cos(\alpha d) + it_{xy}\sin(kd)\sin(\alpha d)\rbrack\nonumber \\
M_{xy} &=&\epsilon_2\lbrack t_{xy}\cos(kd)\cos(\alpha d) - it_{yy}\sin(kd)\sin(\alpha  d)\rbrack \nonumber \\
M_{yx} &=&\epsilon_1 \lbrack it_{xx}\sin(kd)\sin(\alpha d) - t_{xy}\cos(kd)\cos(\alpha d)\rbrack \nonumber. \\
\label{eq:magfield}
\end{eqnarray}
where $\epsilon_1 =  \frac{\epsilon_\perp^{1/2}}{\epsilon_\|^{3/2}}$ and $\epsilon_2= \frac{1}{\sqrt{\epsilon_\perp \epsilon_\|}}$.  The dispersion relations for the two $T$ waves are
the non-trivial solutions to the secular equation formed from  Eqs.~\eqref{eq:disperchiral} and~\eqref{eq:magfield}.

 The most interesting result emerging from Eqs.~\eqref{eq:disperchiral} and~\eqref{eq:magfield} is that the dispersion relations are {\it non-reciprocal}, i.\ e., $\omega(k) \neq \omega(-k)$ in general.  The magnetic field appears only in the off-diagonal elements $t_{xy}$ and $t_{yx}$, which are linear in the field except for very large fields.  The terms involving $t_{xy}$ and $t_{yx}$ in Eq.~\eqref{eq:magfield} are multiplied by $\sin(kd)$ and thus change sign when $k$ changes sign. Thus, the secular equation determining $\omega(k)$ is not even in $k$, implying that the dispersion relations are non-reciprocal.  The non-reciprocal nature of the dispersion relations disappears at $B=0$ even though the off diagonal terms of ${\bf M}(k,\omega)$ are still nonzero, because $\sin(kd)$ appears only to second order.  Also, when the host dielectric is an NLC, the non-reciprocity vanishes because the rotation angle $\alpha =0$ and all terms proportional to $\sin(kd)$ vanish, even at finite $B$.

For a finite {\bf B}, the difference in magnitude  of wave number between a right-moving or left-moving wave is
\begin{equation} \label{eq:delta_k}
\Delta k_i(\omega) = \lvert \mathrm{Re}(k_{i,L})\rvert -\lvert\mathrm{Re}(k_{i,R})\rvert,
\end{equation}
where $i = 1,2$ for the two elliptical polarizations and $ L$ or  $R$ denotes the left-moving or right-moving branch.  If, for example, $\Delta k(\omega) \neq 0$, then the left- and right-moving waves have different magnitudes of  wave numbers for a given frequency and are non-reciprocal. 

\subsection{Faraday Rotation and Ellipticity}

By solving for $k(\omega)$ using either Eqs.~\eqref{eq:nlc_parallel_disp} or~\eqref{eq:rotyz2} for an NLC, or~\eqref{eq:disperchiral} for a CLC,  one finds that the two modes polarized perpendicular to {\bf B} and  propagating along the nanoparticle chain have, in general,  different wave vectors.  For the NLC, we denote these solutions $k_+(\omega)$ and $k_-(\omega)$, while for the CLC, we denote them $k_1(\omega)$ and $k_2(\omega)$. 

We first discuss the case of an NLC host and ${\bf B} \| \hat{z}$. Then the two solutions represent left- and right-circularly polarized waves propagating along the chain.  A linearly polarized mode therefore represents an equal-amplitude mixture of the two circularly polarized modes.  This mixture undergoes a {\it rotation} of the plane of polarization as it propagates down the chain and is analogous to the usual Faraday effect in a {\it bulk} dielectric.   The angle of rotation per unit chain length may be written
\begin{equation}
\label{eq:angle_single}
\theta(\omega) = \frac{1}{2} \left[k_+(\omega) - k_-(\omega)\right].
\end{equation}

\begin{figure}[t]
\includegraphics[width=0.45\textwidth]{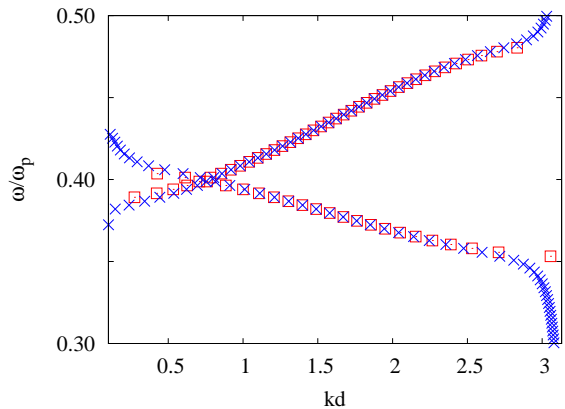} 
\caption{Red open squares: dispersion relations for plasmon waves elliptically polarized in the $yz$ plane and propagating along a chain of nanoparticles described by a Drude dielectric function and assuming no damping. The chain is assumed immersed in an NLC with director parallel to the chain ($\hat{z}$),  ${\bf B} \| \hat{x}$ and $\omega_c/\omega_p = 0.007$.   Blue $x$'s: same as red open squares, but assuming single-particle damping corresponding to $\omega_p\tau = 100$.      For $\omega_p = 1.0\times 10^{13}$ sec$^{-1}$, the chosen $\omega_c/\omega_p$ corresponds to about $2$ Tesla. }  
\label{figure3}
\end{figure}

In the absence of damping, $\theta$ is real.  If $\tau$ is finite, the electrons in each metal or semiconductor particle will experience damping, leading to an exponential decay of the plasmonic waves propagating along the chain. This damping is automatically  included in the above formalism, and can be seen most easily if only nearest neighbor coupling is included. The quantity
\begin{equation}
\theta(\omega) = \theta_1(\omega) + i\theta_2(\omega) 
\label{eq:thetatau}
\end{equation}
is the {\it complex} angle of rotation per unit length of a linearly polarized wave propagating along the chain of  particles.   $\mathrm{Re}\lbrack\theta(\omega)\rbrack$ represents the angle of rotation of a linearly polarized wave (per unit length of chain), and $\mathrm{Im}\lbrack\theta(\omega)\rbrack$ is the corresponding Faraday ellipticity,  i.\ e., the amount by which the initially linearly polarized wave becomes elliptically polarized as it propagates along the chain.  

\begin{figure*}[t]
\centering
\includegraphics[width=0.7\textwidth]{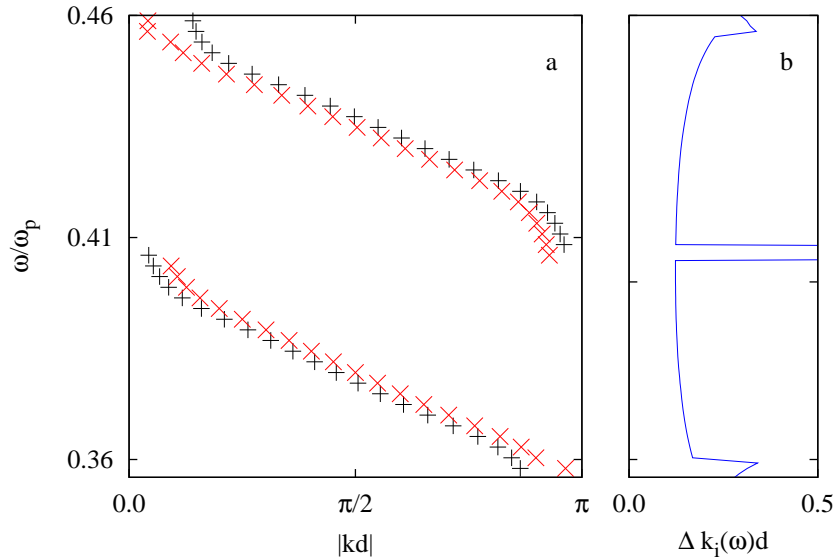} 
\vspace{-0.3in}
\caption{(a). Black $+$ symbols: the two dispersion relations for right-moving transverse plasmon waves propagating along a chain of Drude nanoparticles immersed in a CLC host with ${\bf B}\| \hat{z}$, plotted as a function of $\lvert kd \rvert$. Red x's: same quantities but for left-moving plasmon waves. We assume that $\omega_c/\omega_p = 0.007$, $\omega_p\tau = 100$, and the twist angle $\alpha d = \pi/6$.  (b). The absolute value $|\Delta k_i d|$  of the difference $\Delta k_id = |Re(k_{i,L}d)| - |Re(k_{i,R}d)|$, between the normalized wave vectors for  left-propagating and right-propagating modes of the the two branches, as given by Eq.~\eqref{eq:delta_k}, as functions of frequency.
$L, R$ refer to the left-moving or right-moving waves, and $i$ (i = 1,2) labels the two branches for each direction.   Note that a non-zero value of $|\Delta k_i(\omega)d|$ implies that for a given frequency the left- and right-traveling waves have different wave vectors.  Numerically, we find that $|\Delta k_i(\omega)d|$ is independent of  $i$. There is a gap between the two curves in Fig.~\ref{figure4}(b).   This gap corresponds to a region where the waves propagate in a single direction only.  These {\it one-way modes} occur in the upper branch near $|kd| = \pi$, where there is a small region where only the red branch has solutions, and in the lower branch near $|kd| = 0$, where only the black branch has solutions.  One-way wave propagation occurs only in the region between the two horizontal lines in Fig~\ref{figure4}(b). }
\label{figure4}
\end{figure*}
In the case of a CLC host, neither of the two $T$ modes is circularly polarized in general.  Thus, the propagation of a linearly polarized wave along the chain cannot be simply interpreted in terms of Faraday rotation.

\section{Numerical Illustrations}

We now numerically evaluate the dispersion relations presented in the previous section assuming the host is the liquid crystal known as E7.  This liquid crystal was described  by M\"{u}ller~\cite{muller}, from whom we take the dielectric constants $\epsilon_\|$ and $\epsilon_\perp$.  We first consider the case of an a NLC host with both the director and an applied magnetic field parallel to the chain axis $\hat{z}$. To illustrate the predictions of our simple expressions, we take $a/d = 1/3$, and assume a magnetic field such that the ratio $\omega_c/\omega_p = 0.07$ or $0.007$ as indicated in the Figures.  For a typical plasma frequency of $\sim 10^{13}$ sec$^{-1}$, the ratio of $0.007$ would correspond to a magnetic induction of $B \sim 2T$ .  We consider both the undamped and damped cases; in the latter, we choose $\omega_p\tau = 100$.  For propagating waves we choose solutions for which $\mathrm{Im}\lbrack k_\pm\rbrack > 0$ so that these waves decay to zero, as expected, when $z\rightarrow \infty$ when $\mathrm{Re} \  k >0$.

The calculated dispersion relations for the two circular polarizations of plasmonic wave are shown in Fig.~\ref{figure1} with and without single-particle damping.   The splitting between the two circularly polarized $T$ waves is too small to be seen on the scale of this plot.  The difference can be seen through its effect on the Faraday rotation angle, which is shown in Fig.~\ref{figure2} .  In this, and all subsequent plots, we have calculated far more points than are shown in the Figure, so that effectively the entire range $0 < kd < \pi$ is included.

In Fig.~\ref{figure2}, we  plot the corresponding quantity $\theta(k)d$, the rotation angle for a distance equal to one interparticle spacing.   When there is no damping, we find that the real part of $\theta d$ is very small and that the imaginary part is zero.  Both become larger when damping is included, as we do here by setting  $\omega_p\tau = 100$.  Even in this case,  neither $\mathrm{Re}\lbrack\theta(\omega)d\rbrack$ nor $\mathrm{Im}\lbrack\theta(\omega)d\rbrack$  exceed about $0.005$ radians, showing that a linear incident wave  is rotated only slightly over a single particle spacing (by about 1/4 degree per interparticle spacing for the chosen parameters).   If we assume that the wave {\it intensity} has an exponential decay length of no more than around 20 interparticle spacings in realistic chains, the likely Faraday rotation of such a wave will only be 3-4 degrees over this distance.  The present numerical calculations also suggest that $\theta(k)d$ is very nearly linear in $B$ for a given $k$, so a larger rotation could be attained by increasing $B$; it can also be increased if the electron density is reduced. 

For a chain of Drude particles in a NLC where $B \perp \hat{z}$, we find, using the same parameters and requirements as the previous case, that the two non-degenerate waves (one an $L$ and the other a $T$ wave) become mixed  when $B\neq 0$.  The dispersion relations, again with and without damping, are plotted in Fig.~\ref{figure3}.  When compared to previous work in Ref.~\citenum{pike2013} without the presence of damping, the dispersion relations in Fig.~\ref{figure3} are modified because of the finite damping and presence of the magnetic field.  Decreasing the electron density of the metal or semiconductor at fixed {\bf B} increases the interaction of the coupled $L$ and $T$ mode near their crossing point $kd = 0.7$, although this is not visible in the Figure, for the chosen parameters.

We find that the effect of the magnetic field is such that the two dispersion relations appear to be "repelled" near their crossing point, although this is again not visible in the Figure for the magnetic field considered.  The band gap that opens between the two bands is proportional to the magnetic field.   These features are shown analytically in the Appendix.

Finally, we discuss the case of a chain parallel to the $z$ axis, subjected to a magnetic field along the $z$ axis, and immersed in a CLC whose twist axis is also parallel to $\hat{z}$.  Using the same host dielectric constants given above and a twist angle of $\alpha d = \pi/6$,  we show in Fig.~\ref{figure4}(a) the resulting dispersion relations, i.\ e.,  $\omega/\omega_p$ plotted against $\lvert kd \rvert$, for the two transverse branches.  In particular, we show both transverse branches for a right-moving wave, displayed as black (+)  symbols, and a left-moving wave, displayed as red ($x$) symbols, giving a total of $4$ plots shown in Fig.~\ref{figure4}(a).   The separation between the two $T$ branches is on the order of $0.05 \  \omega/\omega_p$ for all $k$.

In Fig.~\ref{figure4}(b), we plot the corresponding difference in wave number between the left- and right-moving waves as $\Delta k_i(\omega)d$. Since $\Delta k_i(\omega)d$ is nonzero in a wide frequency range, the wave propagation is indeed non-reciprocal in this range.  One-way wave propagation clearly does occur in part of this  range.  Such propagation occurs when, at particular frequencies, waves can propagate only in one of the two directions.   From Fig. ~\ref{figure4}(a), we can see that for the upper dispersion relation, only the right-hand-moving wave can propagate near $kd = \pi$, whereas for the lower one, only the left-hand-moving wave propagates near $kd = 0$.  Thus, there is a gap in the plot of $\Delta k_i(\omega)d$ near $\omega/\omega_p = 0.41$, within which there is only one-way wave propagation.  In Fig.~\ref{figure4}(b), the boundaries of the frequency band for one-way propagation are indicated by the two horizontal lines.
 
\section{Discussion}
The present numerical calculations omit several potentially important factors which could alter the numerical results. The first of these are the effects of particles beyond the nearest neighbors on the dispersion relations~\cite{brong}.  We believe that these further neighbors will mainly modify the details of the dispersion relations without changing the qualititative features
introduced by the magnetic field and the NLC or CLC host.   Another omitted factor is the (possibly large) influence of the particles in disrupting the director orientation of the liquid crystalline host~\cite{lubensky,poulin,stark,kamien,allender}, whether NLC or CLC.   This could be quite important in modifying the dielectric properties of the host liquid near the particles, and could
also cause the positions of the particles themselves to be disturbed, depending on whether they are somehow held in place.   Even though these effects could be quite substantial, we believe that the qualitative effects found in the present calculations, notably the regime of one-way propagation found for certain frequencies in a CLC host, should still be present. We hope to investigate these effects in future work.  Finally, it is known that radiative damping~\cite{weber04}, not included in the quasistatic approximation, can have a significant effect  on the dispersion relations at special frequencies.  Once again, however,
we believe that the qualitative effects discussed in this paper should still be present even if radiation damping is included.    Thus, we believe that  our calculations  do qualitatively describe the combined effects of a liquid crystalline host and an applied magnetic field on  the surface plasmon dispersion relations.

 It should be noted that the magnetic field effects described in this paper are numerically small, for the parameters investigated.  The smallness is caused mainly by the small value of the ratio $\omega_c/\omega_p$, taken here as $0.07$ or $0.007$ depending on the electron density used in the calculation.   To increase this ratio, one could either increase $\omega_c$ (by raising the magnetic field strength), or decrease $\omega_p$ (by reducing the free carrier density in the particle). For the case of a particle chain in a CLC host, any change which increases $\omega_c/\omega_p$ will increases $\Delta k$, leading to a broader frequency rage for one-way wave propagation.

In summary, we have calculated the dispersion relations for plasmonic waves propagating along a chain of semiconducting  or metallic nanoparticles immersed in a liquid crystal and subjected to an applied magnetic field.   For a magnetic field parallel to the chain and director axis of the NLC, a linearly polarized wave is Faraday-rotated by an amount proportional to the magnetic field strength.   For a CLC host and a magnetic field parallel to the chain, the transverse wave solutions become non-reciprocal (left- and right-traveling waves having different dispersion relations) and there are be frequency ranges in which waves propagate only in one direction.   Thus, plasmonic wave propagation can be tuned, either by a liquid crystalline host or a magnetic field, or both. In the future, it may be possible to detect some of these effects in experiments, and to use some of the predicted properties for applications, e. g., in optical circuit design.

\section{Acknowledgments}

This work was supported, in part,  by the Center for Emerging Materials at The Ohio State University, an NSF MRSEC (Grant No.\ DMR-1420451). In addition, this work was supported by the Belgian Fonds National de la Recherche Scientifique FNRS under grant number PDR T.1077.15-1/7.

\section{Appendix}
In this Appendix, we show that the two bands shown in Fig.~\ref{figure3}, which in zero magnetic field cross near $kd = 0.7$, are ``repelled'' in a finite magnetic field ${\bf B} \| {\hat x}$, by an amount proportional to $|{\bf B}|$.   That is, a gap opens up at the crossing point which is proportional  to$|{\bf B}|$.  

The dispersion relations for the coupled $y$ and $z$ polarized waves are obtained from Eqn.~\eqref{eq:rotyz2}.   They have non-trivial solutions when the determinant of the matrix of coefficients vanishes, i.e.,
\begin{eqnarray}
&\left[1+\frac{2}{3}\frac{a^3}{d^3}\frac{\epsilon_\|}{\epsilon_\perp^2}t_{yy}(\omega)\cos(kd)\right]\left[1-\frac{4}{3}\frac{a^3}{d^3}\frac{1}{\epsilon_\perp}
t_{zz}(\omega)\cos(kd)\right] \nonumber \\ &-\frac{8}{9}\frac{a^6}{d^6}\frac{\epsilon_\|}{\epsilon_\perp^3}t_{yz}^2(\omega)\cos^2(kd)= 0.
\label{eq:determ}
\end{eqnarray}

We first consider the case of zero magnetic field. In this case, the off-diagonal components of the t-matrix, namely $t_{yz} = -t_{zy}$, both vanish.  The dispersion relations are then given by 
\begin{equation}
F_1(k, \omega) \equiv 1+\frac{2}{3}\frac{a^3}{d^3}\frac{\epsilon_\|}{\epsilon_\perp^2}t_{yy}(\omega)\cos(kd) = 0
\end{equation}
and
\begin{equation}
F_2(k,\omega) \equiv 1-\frac{4}{3}\frac{a^3}{d^3}\frac{1}{\epsilon_\perp}t_{zz}(\omega)\cos(kd) = 0.
\end{equation}
The two bands will be degenerate when $F_1(k, \omega) = F_2(k, \omega)$, or equivalently
\begin{equation}
\frac{\epsilon_\|}{\epsilon_\perp^2}t_{yy}(\omega) + \frac{1}{\epsilon_\perp}t_{zz}(\omega) = 0.
\label{eq:degenw}
\end{equation} 
Eq.~\eqref{eq:degenw} gives the frequency of the degeneracy, which we denote $\omega_0$.  The corresponding wave vector $k_0$ of the degeneracy is determined by either
\begin{equation}
F_1(k_0, \omega_0) = 0
\end{equation}
or
\begin{equation}
F_2(k_0,\omega_0) = 0.
\end{equation}

Now we consider Eq.~\eqref{eq:determ} with non-zero magnetic field, i.\ e. finite $t_{yz}(\omega)$.   For $k = k_0$, assuming that the band energies $\omega$ are close to their zero-field value $\omega_0$,  we can expand $F_1$ and $F_2$ in Taylor series as 
$F_i(k_0\omega) \sim (\omega - \omega_0) F_i^\prime(k_0, \omega_0)$ for i =1, 2, where
$F_i^\prime(k_0,\omega_0) = [\partial F_i(k_0,\omega)/\partial \omega)]_{\omega=\omega_0}$.
Again to lowest order in B, we can write
$t_{yz}(\omega)\sim t_{yz}(\omega_0)$.  Then the solutions to Eq.~\eqref{eq:determ} are given by
\begin{equation}
F_1^\prime(k_0,\omega_0)F_2^\prime(k_0, \omega_0)(\omega-\omega_0)^2 = \frac{8}{9}\frac{a^6}{d^6}\frac{\epsilon_\|}{\epsilon_\perp^3}t_{yz}^2(\omega_0)\cos^2(k_0d)
\end{equation}
or
\begin{eqnarray}
& |\omega(B)-\omega_0| =  \\
& \pm \{\frac{8}{9}\frac{a^6}{d^6}\frac{\epsilon_\|}{\epsilon_\perp^3}t_{yz}^2(\omega_0)\cos^2(k_0d)/
[F_1^\prime(k_0,\omega_0)F_2^\prime(k_0,\omega_0]\}^{1/2}  \nonumber
\label{eq:finiteb}
\end{eqnarray}
Here $\omega(B)$ represents one of the two band energies at $k = k_0$.  Since $t_{yz}(\omega_0)$ is proportional to B (see below), Eq.~\eqref{eq:finiteb} shows at the splitting between these two band energies at $k = k_0$ is proportional to ${\bf B}$. 

To show that $t_{yz}(\omega_0) \propto B$, we can calculate $t_{yz}$ (and the other components  of  {\bf t}) from Eqs.~\eqref{eq:epsw},~\eqref{eq:aw}, and~\eqref{eq:tmatrix}.  The result, to lowest order in $\delta\epsilon_{yz}(\omega)$ is
\begin{equation}
t_{yz}(\omega) = \frac{\delta\epsilon_{yz}(\omega)}{(1-\Gamma_{yy}\delta\epsilon_{yy}(\omega))(1-\Gamma_{zz}\delta\epsilon_{zz}(\omega))}.
\end{equation}
Since $\delta\epsilon_{yz}(\omega) \propto A(\omega)$, we see that $t_{yz}(\omega) \propto \omega_c \propto B$.  Hence, the splitting
between the two bands at $k = k+0$ is proportional to $|{\bf B}|$.  For the magnitude of B considered in Fig.~\ref{figure3}, this splitting is not visible on the scale of the Figure, but we have tentatively verified numerically that this splitting is present for finite magnetic field.

\end{document}